\documentclass[10pt]{iopart}
\usepackage{graphicx,amssymb,amsfonts,latexsym,color,dcolumn,bm} 
\usepackage[hidelinks]{hyperref}
\usepackage{verbatim}
\newcommand{\ket}[1]{\left|{#1}\right\rangle}
\newcommand{\bra}[1]{\left\langle {#1}\right|}

\newcommand{\req}[1]{Eq.~(\ref{#1})}

\newcommand{\dg}{^{\dagger}}

\def\*#1{\mathbf{#1}}

\begin{document}

\large{This is an author-created, un-copyedited version of an article accepted for publication in the New Journal of Physics as a Fast Track Communication. IOP Publishing Ltd is not responsible for any errors or omissions int his version of the manuscript or any version derived from it. This is the preprint version, the Version of Record is available online at  \href{http://dx.doi.org/10.1088/1367-2630/18/5/052002}{\color{blue}  \ttfamily http://dx.doi.org/10.1088/1367-2630/18/5/052002}.
\normalsize

\thispagestyle{empty}

\clearpage
\setcounter{page}{1}

\title{Microscopic origin of subthermal magnons and the spin Seebeck effect}

\author{I. Diniz and A. T. Costa}
\address{Instituto de F\'isica, Universidade Federal Fluminense, 24210-346 Niter\'oi, RJ, Brazil}
\ead{antc@if.uff.br}

\date{\today}

\begin{abstract}
Recent experimental evidence points to low-energy magnons as the primary contributors to the spin Seebeck effect. This spectral dependence is puzzling since it is not observed on other thermocurrents in the same material. Here, we argue that the physical origin of this behavior is the magnon-magnon scattering mediated by phonons, in a process which conserves the number of magnons. To assess the importance and features of this kind of scattering, we derive the effective magnon-phonon 
interaction from a microscopic model, including band energy, a screened electron-electron interaction and the electron-phonon interaction. Unlike higher order magnon-only scattering, we find that the coupling with phonons induce a scattering which is very small for low-energy (or \textit{subthermal}) magnons but increases sharply above a certain energy  -- rendering magnons above this energy poor spin-current transporters.  
\end{abstract}



\submitto{Fast Track Communication} 


\maketitle

When a magnetic material is subject to a temperature gradient it can inject spin current into an 
attached nonmagnetic metal.  The spin current is detected indirectly; a voltage is created by the 
inverse spin-Hall effect in the nonmagnetic metallic layer (typically Pt) in contact with the 
magnetic material. This phenomenon has been termed spin Seebeck effect (SSE) in analogy with the 
conventional Seebeck effect and it drew much attention as a new way to generate and detect spin 
currents, a central topic in spintronics \cite{BauerSpincaloritronics}.

The SSE was first observed with a metal as the magnetic material \cite{Uchida2008}, but it soon became 
clear that the magnetic material could also be a semiconductor or an 
insulator \cite{SSEsemiconductor,MyersPRB2015,UchidaPRB2015,UchidaPRB2013,Klaui}. This suggests that magnons, 
rather than conduction electrons, are the primary source of SSE. In fact, as insulators do not display charge 
excitations, they provide a much cleaner platform to study the magnon-phonon interactions, which are at the core 
of the phenomenon. In this context, yttriumn iron garnet (YIG) \cite{YIGreview} emerged as a particularly well 
suited material to study the SSE -- with high Curie temperature and large band gap this material remains 
magnetically ordered and electrically insulating from cryogenic to room temperatures.

A wealth of experimental studies on YIG \cite{MyersPRB2015,UchidaPRB2015,UchidaPRB2013,Klaui} revealed 
that the SSE signal displays a strong spectral dependence, i.e. low-energy magnons contribute 
disproportionately to the observed spin current. This particular feature eludes the available 
theoretical models, \cite{AdachiSSEreview,RezendePRB2014} for instance, even if these models have 
successfully described the general properties of the SSE. Here we use a  microscopic model to show 
that the process of magnon-magnon scattering mediated by phonons displays the same strong spectral 
dependence. The  spin current generated in the magnetic material inherits this dependence, 
which ultimately appears in the SSE signal. We argue that the models available so far cannot 
describe this spectral dependence precisely because they lack a microscopic description of the 
magnon momentum relaxation.

Although the SSE was first detected in the transverse configuration, where the spin current is 
perpendicular to the temperature gradient, the longitudinal configuration is much more frequent 
in recent experiments \cite{MyersPRB2015,UchidaPRB2015,UchidaPRB2013,Klaui}. The most striking 
evidence of strong spectral dependence is given precisely in the study of the longitudinal spin 
Seebeck effect (LSSE) as a function of temperature and applied magnetic field. 
The authors of \cite{MyersPRB2015,UchidaPRB2015,UchidaPRB2013} found that the LSSE voltage in a 
Pt/YIG junction is suppressed when a static magnetic field is applied. The suppression behavior 
is remarkable: at temperatures below 35 K the suppression increases abruptly as the sample is 
cooled \cite{MyersPRB2015,UchidaPRB2015}, and above 35 K it saturates but remains large ($\geq$ 20$\%$) 
even at room temperature (see figure 4(a) of \cite{MyersPRB2015}).

As any thermocurrent, we expect the LSSE to be linked to the thermal population of the relevant excitations, 
magnons in this case, and a diffusion length. The highest magnetic field applied in \cite{MyersPRB2015} (70 kOe) 
induces a gap of 9 K  in the magnon spectrum, which is not expected to change appreciably the magnon population 
at room temperature. We are led to consider that the diffusion length must somehow change in the presence of this gap. 
This is precisely what we see in our calculations of the phonon mediated magnon scattering -- even if the gap is relatively 
small, it is sufficiently large to shift the magnon dispersion relation with respect to the phonons, leading to drastically 
different scattering rates (see figure \ref{fig:2}). Direct evidence for the magnetic field dependence of magnon diffusion 
has become available recently: the authors of \cite{LudoYIGlength} found that the magnon diffusion length in YIG reduced 
from 9.6 $\pm$ 1.2 $\mu$m at 10 mT to 4.2 $\pm$ 0.6 $\mu$m at 3.5T at room temperature.  Previous studies \cite{RezendePRB2014} 
argued that the magnon diffusion length is dominated by high order magnon-only scattering; 
if this is the case one could not explain the large change in the diffusion length induced by the magnetic field.

The critical suppression of the LSSE signal at low temperatures (see figure 4 of \cite{MyersPRB2015} 
and figure 5 of \cite{UchidaPRB2015}) provides direct evidence that low energy magnons are the primary 
sources of the spin Seebeck signal. We call such magnons \emph{subthermal}, corresponding to magnons 
bellow $\sim 35$ K in the typical Pt/YIG setup. As one increases the applied magnetic field at a fixed 
temperature $T$, the energy of a magnon mode can be shifted to above $k_B T$, and hence its population 
reduces exponentially as prescribed by the Bose-Einstein statistics. If mainly \emph{subthermal} magnons 
contribute to the SSE signal we expect the SSE suppression to be stable as temperature decreases from room 
temperature until it reaches the \emph{subthermal} magnons region where the suppression increases very rapidly. 
This is precisely the experimental observations and can be understood in the light of our results: the large 
magnon-phonon scattering experienced by long-wavelength magnons makes them poor spin Seebeck generators, 
while for \emph{subthermal} magnons the scaterring is much smaller.

Another evidence relating the thermospin properties with a characteristic length is the measurement of LSSE 
with YIG films of various thicknesses. The authors of \cite{Klaui} have shown that the SSE signal increases 
gradually as thickness grows and saturates above a critical temperature-depend thickness.


In the longitudinal spin Seebeck theory developed in \cite{RezendePRB2014}, based on the Boltzmann approach for 
the magnon spin-current of \cite{ZhangPRL2012}, two kinds of magnon lifetimes are relevant. The first is $\tau_m$, 
related to scattering events which conserve the number of magnons while changing magnon momenta. The second is 
$\tau_{th}$, related to processes which do not conserve the number of magnons. More specifically, 
Zhang \cite{ZhangPRL2012} showed that the magnon diffusion is proportional to $\sqrt{\tau_m \tau_{th}}$ 
while Rezende et al \cite{RezendePRB2014} related this diffusion with magnon accumulation near the Pt layer and 
consequent spin current injection leading to a spin Seebeck signal also proportional to  $\sqrt{\tau_m \tau_{th}}$. 
One can safely assume that the change in SSE with an applied magnetic field is a consequence of the variation of 
$\tau_m$ while $\tau_{th}$ stays almost constant. This comes from the fact that the magnon contribution to 
the thermal conductivity in YIG, which is a process mostly affected by $\tau_{th}$, is insensitive to the applied 
magnetic field (the conductivity of YIG at 300 K changes about $1\%$ with the the application of a 
70 kOe magnetic field \cite{MyersPRB2015}). 


The authors of both \cite{RezendePRB2014} and \cite{ZhangPRL2012} considered that $\tau_m$ should not be 
affected by processes involving phonons, but this turns out to be unjustified. The magnon-phonon processes 
described in our work cannot change the number of magnons and are in fact the main contributors to $\tau_m$. 

To evaluate the relaxation of magnon momentum -- the inverse of $\tau_m$ -- we employ the effective magnon Hamiltonian 
procedure, introduced by Krompiewski and Morkowski \cite{Morkowski197}. The electron-phonon interaction \cite{eph} is added 
to a single-band Hubbard hamiltonian. From this microscopic model, a magnon hamiltonian is derived that includes phonon-mediated magnon scattering. We choose parameters\footnote{On-site interaction U=4.62eV, hopping t=420meV.} for the single-band Hubbard model that describe a strong ferromagnet displaying magnetic excitations analogous to the ones in YIG. A large exchange splitting is present, avoiding any magnon relaxation due to the decay into Stoner excitations \cite{YosidaBook}. Such a simple model should embody the essential physics while allowing a microscopic discussion of the general properties of the magnon-phonon coupling. In this spirit we select representative parameters, rather than attempting to fit a specific system.

\begin{figure}[ht!]
\begin{center}
\includegraphics[width=8.5cm,height=7cm]{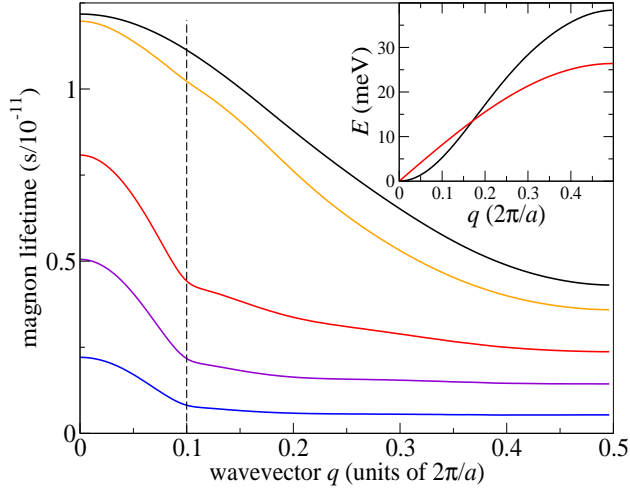} 
\caption{(Color online) Lifetime of magnons ($\tau_m$) in YIG as a function of the normalized wave number; 
only magnon-phonon processes described in \req{eq:H_m-ph} are taken into account. The curves correspond to 
the temperatures 300 K, 150 K, 100 K, 50 K and 1 K, from bottom to top. Regions of low and high relaxation 
can be distinguished; the vertical dotted line indicates the transition to \emph{subthermal} magnons. 
The inset shows the dispersion relations for both magnons (black) and phonons (red); the phonon spectrum corresponds to a Debye temperature of 530K.} 
\label{fig:1}
\end{center}
\end{figure}

Starting from a strongly ferromagnetic ground state $\ket{0}$, the operator 
\begin{equation}
 \beta_{\boldsymbol q}\dg = \sum_{\boldsymbol k} b_{\boldsymbol k +\boldsymbol q,\boldsymbol k} \; a\dg_{\boldsymbol k +\boldsymbol q,\uparrow} a_{\boldsymbol k,\downarrow}  \, 
\end{equation}
generates, in the random-phase-approximation (RPA), a magnon of momentum $\boldsymbol q$ and energy $E_{\boldsymbol q}$. 
The standard notation is used: $a_{\boldsymbol k,\sigma}\dg$ denotes the creation operators for electrons of spin $\sigma$ 
and wave vector $\boldsymbol k$. The coefficients $b_{\boldsymbol k+\boldsymbol q,\boldsymbol k}$ become completely defined 
as we impose canonical commutation relation for the $\beta_{\boldsymbol q}$ operators:
\begin{equation} \label{eq:beta_qcomm}
[\beta_{\boldsymbol q},\beta_{\boldsymbol p}\dg]_{RPA} = \delta_{\boldsymbol q- \boldsymbol p} \, .
\end{equation}
Up to this point the model is equivalent to the usual approach of calculating the dynamical transverse 
spin susceptibility in RPA \cite{overlayerSW}.

The Coulomb interaction will couple electrons with the local charge buildup created by longitudinal phonons. 
The deviation from the equilibrium position of ions can be expanded in terms of the vibrational normal modes 
(phonons) to give the electron-phonon Hamiltonian \cite{eph}
\begin{equation} \label{eq:H_elph}
H_\mathrm{el-ph} = \sum_{\boldsymbol k, \boldsymbol \lambda,\sigma} g_{\boldsymbol \lambda} \; a_{\boldsymbol k,\sigma}\dg a_{\boldsymbol k- \boldsymbol \lambda,\sigma} (b_{\boldsymbol \lambda} + b_{- \boldsymbol \lambda}\dg) \, ,
\end{equation}
where $b_{\boldsymbol \lambda}\dg$ denotes the creation operator for the phonon of wave vector $\boldsymbol \lambda$. 
To reach our goal of describing the phonon mediated magnon-magnon scattering, we search for an effective magnon Hamiltonian 
which should be equivalent to the original Hamiltonian for low-energy excitations. Following \cite{Morkowski197}, the general 
expression for the magnon-phonon Hamiltonian equivalent to \req{eq:H_elph} neglecting higher order magnon interaction is given by
\begin{equation} \label{eq:H_m-ph}
 H_\mathrm{m-ph} = \sum_{\boldsymbol q, \boldsymbol \lambda} \tilde{g}_{\boldsymbol \lambda, \boldsymbol q}  \; \beta_{\boldsymbol q+ \boldsymbol\lambda}\dg \beta_{\boldsymbol q} (b_{\boldsymbol \lambda} + b_{- \boldsymbol \lambda}\dg) \, ,
\end{equation}
where the coefficients $\tilde{g}_{\boldsymbol \lambda, \boldsymbol q}$ have yet to be determined. We can express $\tilde{g}_{\boldsymbol \lambda, \boldsymbol q}$ as a commutator of $H_\mathrm{m-ph}$
\begin{equation} \label{eq:tildeg}
 \tilde{g}_{\boldsymbol \lambda, \boldsymbol q} = [[ \beta_{\boldsymbol q+ \boldsymbol \lambda} , [H_\mathrm{m-ph}, b_{\boldsymbol \lambda}\dg] ] , \beta_{\boldsymbol \lambda}\dg]_{RPA} \, ,
\end{equation}
this is readily obtained inserting equation \req{eq:H_m-ph} and using the commutation relations expressed in \req{eq:beta_qcomm}. If $H_\mathrm{m-ph}$ is to be equivalent to the original Hamiltonian $H_\mathrm{el-ph}$, we can replace the original 
Hamiltonian for $H_\mathrm{m-ph}$ in the commutator in \req{eq:tildeg} to find
\begin{eqnarray}
&\frac{ \tilde{g}_{\boldsymbol \lambda, \boldsymbol q} } { g_{\boldsymbol \lambda} } = \sum_{\boldsymbol k,\sigma} \bra{0} [[ \beta_{\boldsymbol q+\boldsymbol \lambda} , a_{\boldsymbol k,\sigma}\dg a_{\boldsymbol k-\boldsymbol \lambda,\sigma} ] , \beta_{\boldsymbol \lambda}\dg] \ket{0}  \\
&= \sum_{\boldsymbol{k} } \, b_{\boldsymbol q+ \boldsymbol k, \boldsymbol k} (b_{\boldsymbol q + \boldsymbol k + \boldsymbol \lambda, \boldsymbol k}^* - b_{\boldsymbol q + \boldsymbol k, \boldsymbol k - \boldsymbol \lambda}^*) (f_{\boldsymbol k,\downarrow}  - f_{\boldsymbol q + \boldsymbol k + \boldsymbol \lambda,\uparrow} ) \nonumber
\end{eqnarray}
where $f_{\boldsymbol k,\sigma}=  \bra{0} a_{\boldsymbol k,\sigma}\dg a_{\boldsymbol k,\sigma} \ket{0} $. 
Neglecting magnon interactions of higher order, the total Hamiltonian of the magnon-phonon system is 
$H=H_\mathrm{m-ph}+ H_\mathrm{m} + H_\mathrm{ph}$, where the energy of free magnons ($H_\mathrm{m}$) 
and free phonons ($H_\mathrm{ph}$) are defined in the usual way and plotted in the inset of figure \ref{fig:1}. 
Phonons interact directly with the external thermal baths and as such tend to thermalize much faster. 
The effect on the magnon system caused by the interaction with the thermalized phonons can be described 
by a Lindbladian Master equation\cite{MasterEq}. It is then straightforward to evaluate relaxation rates 
as a function of the magnon wave vector $\boldsymbol q$, in the near-equilibrium situation:
\begin{eqnarray} \label{eq:d_p}
\frac{d_{\boldsymbol q}}{2 \pi} &= \sum_{\boldsymbol \lambda} | \tilde{g}_{\boldsymbol \lambda, \boldsymbol q} |^2 [ (1 + N_{\boldsymbol q+ \boldsymbol \lambda} + n_{\boldsymbol \lambda})
\delta(E_{\boldsymbol q+ \boldsymbol \lambda}-E_{\boldsymbol q} + \Omega_{\boldsymbol \lambda}) + \\ 
&+ (n_{\boldsymbol \lambda} - N_{\boldsymbol q+ \boldsymbol \lambda}) \delta(E_{\boldsymbol q+ \boldsymbol \lambda}-E_{\boldsymbol q} - \Omega_{\boldsymbol \lambda})  ]  \, , \nonumber
\end{eqnarray}
where $\Omega_{\boldsymbol \lambda}$ is the frequency of phonons of momentum $\boldsymbol \lambda$, and $N_{\boldsymbol q}$ 
($n_{\boldsymbol q}$) is the equilibrium population of magnons (phonons) of momentum $\boldsymbol q$. 
Both $N_{\boldsymbol q}$ an $n_{\boldsymbol q}$ display temperature dependence which will be inherited by $d_{\boldsymbol q}$.

Figure \ref{fig:1} depicts the surprising behavior of \req{eq:d_p} for a strong magnet such as the YIG: regions of low and high 
relaxation appear naturally in our model. As data for the electron-phonon coupling interaction in YIG are scarce, 
we set $g_{\boldsymbol \lambda} = g_0 ( {|\boldsymbol  \lambda| } / { | \boldsymbol \lambda_{\max} | } )^{1/2}$ which
corresponds to approximating the Bloch states by planes waves \cite{eph}. To make sure the spectral dependence we find 
for the magnon lifetimes is an intrinsic characteristic of the process and not the consequence of a particular 
$g_{\boldsymbol \lambda}$, we plot in figure \ref{fig:2} the lifetimes obtained with a completely flat coupling constant 
$g_{\boldsymbol \lambda} = g_0$. Note that as each magnon mode interacts with a broad range of other modes, the results are 
not sensitive to small variations in the $\boldsymbol \lambda$-dependence of $g_{\boldsymbol \lambda}$. Typical values of 
$g_{\boldsymbol \lambda}$ range from 100 meV to 1 eV \cite{ephConstant}. In the present work we take $g_0=100$meV as a 
representative value -- note that the relaxation rates (Eq. \ref{eq:d_p}) simply scale quadratically with this parameter.

\begin{figure}[ht!]
\begin{center}
\includegraphics[width=9.5cm,height=7cm]{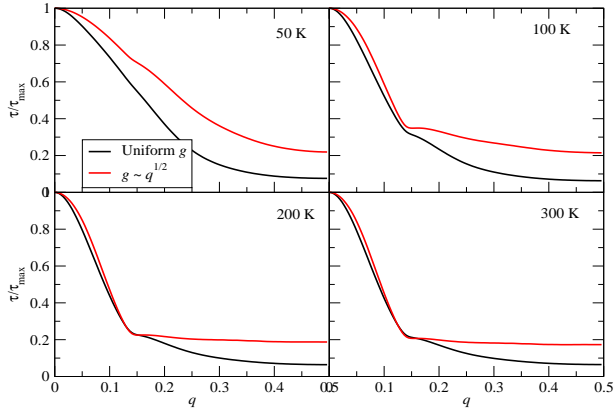}
\caption{(Color online) Comparison of magnons lifetime ($\tau_m$) for different electron-phonon coupling functions. 
Lifetimes are plotted as a function of magnon momentum for a different temperature in each subplot. The general 
results are robust to this change: regions of low relaxation (\emph{subthermal} magnons) and high relaxation 
can always be distinguished.}  
\label{fig:2}
\end{center}
\end{figure}

In adopting the RPA in the effective Hamiltonian of \req{eq:H_m-ph} we neglect the effects of magnon-magnon interactions. 
Nevertheless, we can safely neglect the relaxation given by higher-order processes as we find that the process we describe 
is by far the strongest mechanism, being on average two-orders of magnitude larger than 3-magnon and 4-magnon processes 
described in \cite{RezendePRB2014}. Furthermore, the spectral behavior of the relaxation caused by magnon-magnon processes 
is not in agreement with the evidence of subthermal magnons. For instance, 3-magnon relaxation simply scales linearly with
magnon momentum \cite{3magnonrelax} .

\begin{figure}[ht!]
\begin{center}
\includegraphics[width=8.5cm,height=7cm]{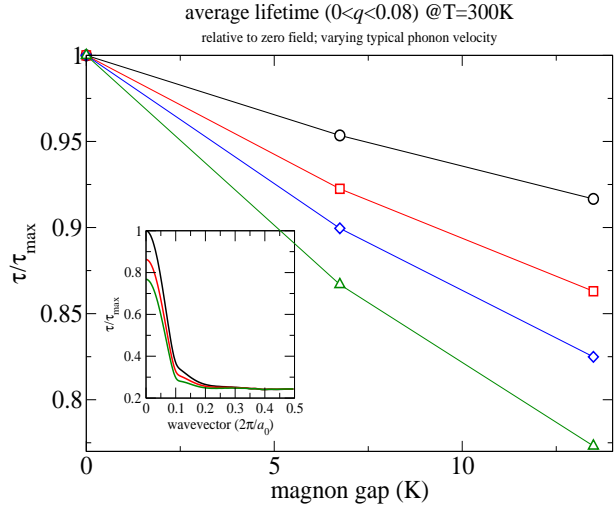}
\caption{(Color online) Ratio of the average of the lifetime with ($\tau$) and without ($\tau_{max}$) external field 
as a function of the field-induced magnon gap at 300 K. The average is performed in the region of \emph{subthermal} magnons. 
The curve is plotted for different Debye temperatures 530 K, 600 K, 660 K and 790 K, from top to bottom. The inset shows the 
spectral dependence of the magnon lifetimes for magnon gaps equal to 0, 7 K and 14 K and Debye temperatures 790 K 
(corresponding to the data-points on the bottom curve of the main plot). Only \emph{subthermal} magnons are appreciably 
affected having higher relaxation for larger gaps.}
\label{fig:3}
\end{center}
\end{figure}

In the presence of an applied static magnetic field the magnon spectrum becomes gapped. Figure~\ref{fig:3} shows how the 
average lifetime of the \emph{subthermal} magnons changes in the presence of a gap of up to 9 K. The inset shows the whole 
spectral dependence: when the magnetic field is applied inducing a gap, \emph{subthermal} magnons get scattered more 
efficiently while the high energy magnons are barely affected. The lifetime suppression will correlate to a suppression 
of the SSE signal. This is because the SSE signal is a function of both the magnon lifetime and the magnon population, 
but at room temperature the magnon population remains essentially the same in the presence of such a small gap. 
Room temperature SSE suppression was measured in YIG/Pt to be about 20$\%$  with a 9 K gap \cite{MyersPRB2015,UchidaPRB2015},  our calculations agree qualitatively with this measurements and show that a considerable  room temperature SSE suppression is a robust phenomenon. 

It has been argued \cite{MyersPRB2015} that the SSE generates a fundamentally different thermocurrent 
as it necessarily involves not just the thermal excitation and transport of magnons but also the spin 
transfer across an interface, which could depend strongly on the magnon energy. The behavior of the spin 
transfer coefficient is yet to be fully understood and it can certainly influence the spectral non-uniformity 
of the SSE; nevertheless our results indicate that the main features can be understood simply as a consequence 
of the spectral dependence of the magnon lifetime. We also point out that the reduction in the magnon diffusion length under applied magnetic field has been directly observed, in the experiment reported in \cite{LudoYIGlength} the spin current is both injected and detected in the YIG layer showing that the effect is present event when there is no top nonmagnetic layer. This is in agreement with experimental results where changing the interface did not affect the SSE  signal suppression: the authors of \cite{UchidaPRB2013} found analogous SSE suppression under magnetic field in both Pt/YIG and Au/YIG samples. 

It is noteworthy that the relaxation of magnon momentum described in the present work might not affect every magnon transport phenomena in the same way.
For instance, thermal conductivity due to magnons in YIG as a function of external field does not saturates as quickly as the SSE signal \cite{YIGthermalconduc}. There is a longstanding discussion concerning the need of the magnon-phonon interaction to describe the behavior of thermal conductivity in YIG. This discussion started in \cite{Walton} and it is still very much alive; see \cite{SergioYIGconduct} and \cite{BoonaYIGconduct} for recent, opposing viewpoints on this subject.

In summary, we used a microscopic model to investigate the magnon relaxation caused by phonon-mediated scattering. 
The spectral dependence of this relaxation and its behavior on the presence of an external magnetic field accounts 
for the features observed in the LSSE signal. Both the room-temperature suppression under high magnetic field and 
the critical suppression at low temperatures appear spontaneously in the model. The simplicity of our model attests 
that this should be present not only in YIG but in any strong magnet in which the magnon momentum relaxation is 
dominated by the phonon mediated magnon scattering. We expect that as more data on the electron-phonon interaction 
becomes available, quantitative calculations in specific systems  will become viable. Finally, our results give a 
clear meaning to the emerging notion of \emph{subthermal} magnons while insisting, as proposed by other authors, 
that different kinds of magnon relaxation are relevant to the Spin Seebeck effect. 

\ack 

The authors acknowledge finacial support from CNPq, CAPES and FAPERJ and gratefully thank R. B. Muniz for useful discussions.


\end{document}